\documentclass[aps,prl,groupedaddress,superscriptaddress,twocolumn,notitlepage,bibnotes]{revtex4-1}

\usepackage{newpxtext,newpxmath}

\let\coloneqq\relax

\usepackage{newpxtext,newpxmath}

\usepackage[utf8]{inputenc}

\usepackage{amsthm}
\usepackage{amssymb}
\usepackage{amsmath}
\usepackage{bbold}
\usepackage{bbm}
\usepackage[pdftex, backref=page]{hyperref}
\usepackage{braket}
\usepackage{dsfont}
\usepackage{mathdots}
\usepackage{mathtools}
\usepackage{enumerate}
\usepackage[shortlabels]{enumitem}
\usepackage{csquotes}
\usepackage{stmaryrd}
\usepackage[cal=boondox]{mathalfa}
\usepackage{graphicx}
\usepackage{stackengine}
\usepackage{scalerel}
\usepackage{tensor}       
\usepackage{array}
\usepackage{makecell}
\newcolumntype{x}[1]{>{\centering\arraybackslash}p{#1}}
\usepackage{tikz}
\usepackage{pgfplots}
\usetikzlibrary{shapes.geometric, shapes.misc, positioning, arrows, arrows.meta, decorations.pathreplacing, decorations.pathmorphing, patterns, angles, quotes, calc}
\usepackage{booktabs}
\usepackage{xfrac}
\usepackage{siunitx}
\usepackage{centernot}
\usepackage{comment}
\usepackage{chngcntr}
\usepackage[justification=justified,format=plain]{caption}
\usepackage{subcaption}
\usepackage{ragged2e}

\newtheorem{thm}{Theorem}
\newtheorem*{thm*}{Theorem}

\newtheorem*{prop*}{Proposition}
\newtheorem{lemma}[thm]{Lemma}
\newtheorem*{lemma*}{Lemma}

\newtheorem*{cor*}{Corollary}

\newtheorem*{cj*}{Conjecture}

\newtheorem*{Def*}{Definition}

\newtheorem*{question*}{Question}

\newtheorem*{problem*}{Problem}

\makeatletter
\def\thmhead@plain#1#2#3{%
  \thmname{#1}\thmnumber{\@ifnotempty{#1}{ }\@upn{#2}}%
  \thmnote{ {\the\thm@notefont#3}}}
\let\thmhead\thmhead@plain
\makeatother

\theoremstyle{definition}

\makeatletter
\newcommand{\manualifempty}[3]{%
  \edef\@tempa{#1}%
  \ifx\@tempa\@empty
    #2
  \else
    #3
  \fi
}
\makeatother

\makeatletter
\newtheoremstyle{manualstyle}
  {3pt}{3pt}{\itshape}{}{\bfseries}{.}{ }{}

\theoremstyle{manualstyle}

\newtheorem{manualthminner}{Theorem}

\newtheorem{manualpropinner}{Proposition}

\newtheorem{manuallemmainner}{Lemma}

\newtheorem{manualcorinner}{Corollary}

\makeatother

\newcommand{\bb}{\begin{equation}\begin{aligned}\hspace{0pt}}
\newcommand{\bbb}{\begin{equation*}\begin{aligned}}
\newcommand{\ee}{\end{aligned}\end{equation}}
\newcommand{\eee}{\end{aligned}\end{equation*}}

\newcommand{\sumno}{\sum\nolimits}

\renewcommand{\epsilon}{\varepsilon}

\newcommand{\tcb}[1]{{\color{blue!80!black} #1}}

\newcommand{\id}{\mathds{1}}

\newcommand{\locc}{\mathrm{LOCC}}

\newcommand{\ppt}{\mathrm{PPT}}

\DeclareMathOperator{\Tr}{Tr}

\DeclareMathAlphabet{\pazocal}{OMS}{zplm}{m}{n}

\newcommand{\lsmatrix}{\left(\begin{smallmatrix}}
\newcommand{\rsmatrix}{\end{smallmatrix}\right)}

\stackMath

\stackMath

\makeatletter
\newcommand*\rel@kern[1]{\kern#1\dimexpr\macc@kerna}
\newcommand*\widebar[1]{%
  \begingroup
  \def\mathaccent##1##2{%
    \rel@kern{0.8}%
    \overline{\rel@kern{-0.8}\macc@nucleus\rel@kern{0.2}}%
    \rel@kern{-0.2}%
  }%
  \macc@depth\@ne
  \let\math@bgroup\@empty \let\math@egroup\macc@set@skewchar
  \mathsurround\z@ \frozen@everymath{\mathgroup\macc@group\relax}%
  \macc@set@skewchar\relax
  \let\mathaccentV\macc@nested@a
  \macc@nested@a\relax111{#1}%
  \endgroup
}

\newcommand{\fakepart}[1]{
 \par\refstepcounter{part}
  \sectionmark{#1}
}

\counterwithin*{equation}{part}
\counterwithin*{thm}{part}
\counterwithin*{figure}{part}

\tikzset{meter/.append style={draw, inner sep=10, rectangle, font=\vphantom{A}, minimum width=30, line width=.8, path picture={\draw[black] ([shift={(.1,.3)}]path picture bounding box.south west) to[bend left=50] ([shift={(-.1,.3)}]path picture bounding box.south east);\draw[black,-latex] ([shift={(0,.1)}]path picture bounding box.south) -- ([shift={(.3,-.1)}]path picture bounding box.north);}}}
\tikzset{roundnode/.append style={circle, draw=black, fill=gray!20, thick, minimum size=10mm}}
\tikzset{squarenode/.style={rectangle, draw=black, fill=none, thick, minimum size=10mm}}

\definecolor{Blues5seq1}{RGB}{239,243,255}
\definecolor{Blues5seq2}{RGB}{189,215,231}
\definecolor{Blues5seq3}{RGB}{107,174,214}
\definecolor{Blues5seq4}{RGB}{49,130,189}
\definecolor{Blues5seq5}{RGB}{8,81,156}

\definecolor{Greens5seq1}{RGB}{237,248,233}
\definecolor{Greens5seq2}{RGB}{186,228,179}
\definecolor{Greens5seq3}{RGB}{116,196,118}
\definecolor{Greens5seq4}{RGB}{49,163,84}
\definecolor{Greens5seq5}{RGB}{0,109,44}

\definecolor{Reds5seq1}{RGB}{254,229,217}
\definecolor{Reds5seq2}{RGB}{252,174,145}
\definecolor{Reds5seq3}{RGB}{251,106,74}
\definecolor{Reds5seq4}{RGB}{222,45,38}
\definecolor{Reds5seq5}{RGB}{165,15,21}

\allowdisplaybreaks

\usepackage[most,breakable]{tcolorbox}
\makeatletter

\def\boxed@gobegin#1{\def\@tempa{#1}\def\@tempb{orange}\ifx\@tempa\@tempb\begin{tcolorbox}[colback=red!15,colframe=orange!70,breakable,enhanced]\else\begin{tcolorbox}[colback=Blues5seq1,colframe=Blues5seq5,breakable,enhanced]\fi}
\def\boxed@gobegin@empty{\begin{tcolorbox}[colback=Blues5seq1,colframe=Blues5seq5,breakable,enhanced]}
\makeatother    
\renewcommand*{\backref}[1]{}
\renewcommand*{\backrefalt}[4]{}

\counterwithin*{equation}{part}
\counterwithin*{thm}{part}
\counterwithin*{figure}{part}
\usepackage{pgfplots}
\pgfplotsset{width=10cm,compat=1.9}
\usetikzlibrary{decorations.markings}

\let\tcb\relax

\begin{document}

\title{Cost of entanglement simplified: a single-letter formula (after all)}

\author{Francesco Anna Mele}
\email{fmele@caltech.edu}
\affiliation{Institute for Quantum Information and Matter, Caltech, Pasadena, CA 91125, USA}

\author{Samrat Sen}
\email{samrat.sen@sns.it}
\affiliation{Scuola Normale Superiore, Piazza dei Cavalieri 7, 56126 Pisa, Italy}

\author{Ludovico Lami}
\email{ludovico.lami@gmail.com}
\affiliation{Scuola Normale Superiore, Piazza dei Cavalieri 7, 56126 Pisa, Italy}

\begin{abstract}
A long-standing goal in quantum Shannon theory is to find a single-letter formula for an entanglement measure that has a strong operational meaning. Here we achieve this goal for the exact entanglement cost under operations with positive partial transpose (PPT), for arbitrary bipartite states. This cost was previously characterised as the \emph{limit} of single-letter semidefinite programs with an increasing number of variables and constraints. Here we prove that this limit is unnecessary: the hierarchy collapses at a finite, explicit level depending only on the dimension. Thus, for every fixed dimension, the exact PPT entanglement cost is given by a 
single-letter (efficiently computable) semidefinite program. To our knowledge, it is the first time that an operationally meaningful entanglement measure is shown to be expressible in this form.

\end{abstract}


\maketitle

\let\oldaddcontentsline\addcontentsline
\renewcommand{\addcontentsline}[3]{}
\fakepart{Main text}

\textbf{\em Introduction.}---
\tcb{Entanglement is one of the most important driving fuels behind several advantages quantum theory offers over its classical counterpart, such as communication~\cite{bennett92,bennett93,bennett96}, communication complexity~\cite{Buhrman2001,Brukner2004}, measurement precision~\cite{Giovannetti2004}, computation~\cite{rauss01} and cryptography~\cite{ekert91}. The physical framework in which entanglement manipulation is best considered is local operations and classical communication (LOCC); in this scenario two or more spatially separated parties perform local operations on their subsystems and communicate classically amongst themselves. Despite the apparent simplicity of this framework, the intricate mathematical structure of LOCC renders precise characterisation of entanglement manipulation notoriously intractable~\cite{Chitambar2014}. For example, in the task of entanglement dilution, the goal is to prepare $n$ copies of a given quantum state $\rho = \rho_{AB}$ with as few two-qubit maximally entangled states as possible. This defines the cost of preparing that state, which in LOCC boils down to evaluating the expression
\bb
E_{c,\locc}(\rho) = \lim_{n \to \infty}\frac1n E_f\big(\rho^{\otimes n}\big)\, ,
\label{eq:locc-cost}
\ee
with $E_f$ being the entanglement of formation~\cite{bennett96,wootters98}. Regularised formulas of this kind pervade entanglement theory~\cite{horoRMP}; they are analytically hard to control and furthermore seldom computable: on the one hand, the dimension of $\rho^{\otimes n}$ is exponential in $n$, so the terms rapidly become infeasible to evaluate; on the other, subadditivity makes every term an upper bound on the limit, while nothing is known about the difference $\frac1n E_f\big(\rho^{\otimes n}\big) - E_{c,\locc}(\rho)$; hence, truncating the sequence yields an upper bound with no guarantee on the approximation error.

Facing LOCC entanglement manipulation head-on is thus undoubtedly difficult, which is why the scientific community has instead adopted the approach of studying entanglement manipulation in larger but more tractable classes that approximate LOCC from outside; doing so yields tight upper bounds on what LOCC can achieve. Operations with positive partial transpose (PPT) form a semidefinite relaxation of LOCC and are one of the best-known of these outer approximations, going back to the seminal work of Rains~\cite{rains99,rains01}. This PPT relaxation provides a simplified model of entanglement
manipulation, which has long been
studied~\cite{Eggeling2001,audenaert2003entanglement,%
IshizakaPlenio2005,MatthewsWinter2008,WangDuan2016,%
WangDuanRains2017,FangWangTomamichelDuan2019,%
RegulaFangWangGu2019,wang2020cost} and still provides physical
insights~\cite{WangDuan2017,LuKuoLin2022,WangJingZhu2025,%
Ganardi2026,ZhuWang2026ExactDist,Lami2026PPTDistillation}.

In this work, we focus on this simplified model of entanglement
manipulation, in particular on the task of entanglement dilution
under PPT operations~\cite{audenaert2003entanglement,WangDuan2017,%
DuanLuoLi2018,wang2020cost,Lami2023,LMR25,WangJingZhu2025,%
Kondra2026StrongIrreversibility}. A channel $\Lambda$ is called PPT if its partial
transposition $\Gamma \circ \Lambda \circ \Gamma$ is also a
valid quantum channel, where $\Gamma$ is the partial transpose
operation~\cite{peres} on $B$, defined as
$\Gamma(X_A \otimes Y_B) = X_A \otimes Y_B^{\mathsf T}$.
In its usual definition, the entanglement cost allows an error,
albeit one that vanishes asymptotically. There is another
quantity, the exact PPT entanglement cost, which is much more
stringent but of considerable mathematical interest: it
requires that the error be \textit{exactly} zero for all
sufficiently large numbers of copies. Figure~\ref{fig:exact-ppt-dilution} illustrates the task.
\begin{figure}[!h]
  {\centering
  \includegraphics[width=\columnwidth]{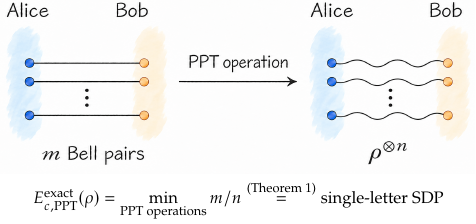}\par}
  \caption[Exact PPT entanglement dilution.]{\protect\justifying
  Exact PPT entanglement dilution. A PPT operation converts
  $m$ Bell pairs shared by Alice and Bob into $n$ exact copies of the
  bipartite state $\rho$. Theorem~\ref{thm:collapse} shows that the
  minimum asymptotic rate $m/n$, which is by definition the \emph{exact PPT entanglement cost} of $\rho$, can be computed by solving a
  single-letter SDP.}
  \label{fig:exact-ppt-dilution}
\end{figure}

The exact PPT entanglement cost was first studied in a
pioneering work by Audenaert, Plenio, and
Eisert~\cite{audenaert2003entanglement}, who showed that
$E_{c,\ppt}^{\mathrm{exact}}(\rho)$ can be evaluated exactly
for all bipartite states that have ``zero
binegativity''~\cite{vidal2002computable,plenio2005logarithmic,Satoshi,werner02}.
Specifically, by denoting as $X^\Gamma$ the partial transpose of a bipartite operator $X$ and as $|X| \coloneqq \sqrt{X^\dagger X}$ its absolute value, then
\bb
\big|\rho^\Gamma\big|^\Gamma \geq 0\quad  \Rightarrow \quad E_{c,\ppt}^{\mathrm{exact}}(\rho)
= \log_2\big\|\rho^\Gamma\big\|_1\, .
\label{eq:zero-binegativity}
\ee
Here, $E_N\coloneqq \log_2\big\|\rho^\Gamma\big\|_1$ is the widely known logarithmic negativity~\cite{vidal2002computable,plenio2005logarithmic},
obtained by simply evaluating the trace norm
$\|\cdot\|_1 \coloneqq \Tr|\cdot|$ of the partially transposed
state. However, not all quantum states possess zero
binegativity, and logarithmic negativity provides, in general,
only a lower bound to the exact cost~\cite{wang2020cost,wang2023exact,WangWildeErrata2024}. Another contribution crucial to this development was that
of Wang and Wilde~\cite{wang2020cost,wang2023exact,WangWildeErrata2024}, who
derived a regularised characterisation of the exact PPT
entanglement cost for all quantum states. The title of the present paper is in fact inspired by the title of Ref.~\cite{wang2020cost}, \emph{Cost of Quantum Entanglement Simplified}, where the authors motivate the study of exact entanglement dilution under PPT operations as an operationally meaningful way to investigate the structure of quantum entanglement in a simplified yet insightful setting. Despite the regularisation involved  in the formula of the cost provided in~\cite{wang2020cost,wang2023exact,WangWildeErrata2024}, some of us recently showed that the it can nevertheless be efficiently approximated
through a converging hierarchy of single-letter
semidefinite programs~\cite{LMR25}. In other words, the
cost can be expressed as the \emph{limit} of single-letter
quantities obtained from semidefinite programs.

While this provided a satisfactory solution for approximating
the cost to arbitrary precision, it remained open whether
the cost could be expressed in terms of a single, single-letter
formula, i.e.\ without taking the \emph{limit} of a sequence
of such expressions. This question is motivated by a
long-standing goal in quantum Shannon theory: finding
efficiently computable, single-letter formulae for entanglement
measures with a strong operational meaning.

In this work, we achieve this goal for the exact PPT
entanglement cost, obtaining an efficiently computable,
single-letter formula valid for arbitrary bipartite states.
More precisely, we prove that the hierarchy of Ref.~\cite{LMR25} collapses at a finite level depending only on the dimension,
uniformly over all states. Consequently, the exact PPT
entanglement cost can be expressed in terms of a finite single-letter semidefinite program. Semidefinite programs are appealing for both numerical and analytical reasons, as they can be numerically solved efficiently with rigorous guarantees and their primal--dual structure often helps mathematically prove properties of the quantities they compute~\cite{vandenberghe1996semidefinite,watrous2009semidefinite}.

We now describe the semidefinite-program hierarchies of Ref.~\cite{LMR25} in more detail. For a density operator $\rho$ and an integer $p \geq 0$, the lower hierarchy is~\cite{LMR25}
\begin{equation}
\label{eq:chi}
\begin{array}{@{}r@{\;}l@{\;}c@{\;}l@{}}
\chi_p(\rho) & \coloneqq &
\displaystyle\min\limits_{\mathclap{S_0,\ldots,S_p}} &
\quad\Tr S_p \\[3pt]
&
\hspace{0.6em}
\scalebox{0.9}{$
\begin{array}{@{}r@{}}
  -S_0\leq \\[-2pt]
  -S_1\leq \\[-2pt]
  \phantom{\vdots} \\[3pt]
  -S_p\leq
\end{array}
$}
&
\scalebox{0.9}{$
\begin{array}{@{}c@{}}
  \rho^\Gamma \\[-2pt]
  S_0^\Gamma \\[-2pt]
  \vdots \\[3pt]
  S_{p-1}^\Gamma
\end{array}
$}
&
\scalebox{0.9}{$
\begin{array}{@{}l@{}}
  \leq S_0 \\[-2pt]
  \leq S_1 \\[-2pt]
  \phantom{\vdots} \\[3pt]
  \leq S_p
\end{array}
$}
\end{array}
\end{equation}
The sequence is nondecreasing, starts at $\chi_0(\rho) = \big\|\rho^\Gamma\big\|_1$, and its limit characterises the exact PPT entanglement cost~\cite{LMR25}:
\bb\label{eq_cost}
E_{c,\ppt}^{\mathrm{exact}}(\rho) = \lim_{p\to\infty} \log_2\chi_p(\rho)\,,
\ee
Specifically, the quantity $\log_2\chi_p(\rho)$ converges to the exact cost exponentially fast in $p$~\cite{LMR25}.
However, nothing in the argument of Ref.~\cite{LMR25} prevents the
sequence $\chi_p(\rho)$ from becoming constant for all sufficiently
large $p$, in which case the cost can be computed by solving a single
semidefinite program. In this work we show that this is the case.
Letting $D$ be the total Hilbert space dimension, our main result
(Theorem~\ref{thm:collapse}) establishes that the cost can be expressed
in terms of the semidefinite program in~\eqref{eq:chi} evaluated at a
finite, explicit level depending only on $D$. More precisely, we prove that
\bb
E_{c,\ppt}^{\mathrm{exact}}(\rho) = \log_2\chi_p(\rho)
\ee
for $p = \lfloor D/2\rfloor$ (and thus for every $p \geq \lfloor D/2\rfloor$). Note that a
dimension dependence is unavoidable, since Brinster et al.~\cite{Brinster2026} showed that no fixed finite level $p$ suffices for all states across all dimensions.

As a by-product of our analysis, we also shed light on the mathematical structure of bipartite states. Ishizaka showed that every two-qubit state has zero
binegativity~\cite{Satoshi}, i.e.\ $\bigl|\rho^\Gamma\bigr|^\Gamma
\geq 0$, so that~\eqref{eq:zero-binegativity} gives the exact PPT
entanglement cost of all two-qubit states. In the Supplemental
Material (Theorem~\ref{thm:greedy}), we prove a higher-dimensional
counterpart of Ishizaka's result: for every bipartite state $\rho$,
\bb\label{eq_ppppppt}
  \Biggl(
    \underbrace{
      \Bigl|\cdots
        \bigl|\lvert\rho^\Gamma\rvert^\Gamma\bigr|^\Gamma
      \cdots\Bigr|
    }_{p\ \text{applications of } X\mapsto\lvert X^\Gamma\rvert}
  \Biggr)^\Gamma \geq 0
\ee
for every integer $p\geq 3D^{3/2}\ln(32D)$, where $D$ is the
total dimension. Thus, starting from any bipartite state,
repeatedly taking the absolute value of the partial transpose
yields a PPT operator in finitely many steps. As an application of this property, the resulting PPT operator in \eqref{eq_ppppppt} yields an upper bound on the exact PPT entanglement cost (as it follows by~\cite[Lemma~4]{Brinster2026}).

\medskip
\textbf{{\em Finite collapse.}} ---  We start with our main result.
\begin{thm}[(Single-letter formula for the exact PPT entanglement cost)]\label{thm:collapse}
For any state $\rho$ on $\mathbb{C}^{d_A}\otimes \mathbb{C}^{d_B}$,  
\bb
E_{c,\ppt}^{\mathrm{exact}}(\rho)=\log_2\chi_{p_\ast}(\rho) \, ,
\label{eq:collapse}
\ee
where $\chi_{p_\ast}(\rho)$ is the (single-letter) semidefinite program in \eqref{eq:chi} of level $p_\ast \coloneqq \left\lfloor d_A d_B/2\right\rfloor$.
\end{thm}


Let us start by understanding why large values of the level $p$ are not needed. The dual program of \eqref{eq:chi} reads
\begin{equation}
\label{eq:dual}
\begin{array}{@{}r@{\;}l@{\;}c@{\;}l@{}}
\chi_p(\rho) & = &
\displaystyle\max\limits_{\mathclap{B_0,\ldots,B_p}} &
\quad\Tr[\rho B_0] \\[3pt]
&
\hspace{0.6em}
\scalebox{0.9}{$
\begin{array}{@{}r@{}}
  -B_1\leq \\[3pt]
  -B_2\leq \\
  \phantom{\vdots} \\[3pt]
  -\id\leq
\end{array}
$}
&
\scalebox{0.9}{$
\begin{array}{@{}c@{}}
  B_0^\Gamma \\[3pt]
  B_1^\Gamma \\
  \vdots \\[3pt]
  B_p^\Gamma
\end{array}
$}
&
\scalebox{0.9}{$
\begin{array}{@{}l@{}}
  \leq B_1 \\[3pt]
  \leq B_2 \\
  \phantom{\vdots} \\[3pt]
  \leq \id
\end{array}
$}
\end{array}
\end{equation}
We regard a feasible sequence as a path from its initial witness $B_0$ to the identity $\id$. The operators $B_j$ are positive for $j \geq 1$, and their traces can only increase, because $S \geq \pm H$ implies $\Tr[S] \geq \|H\|_1$. The constant identity path has objective one, so an optimal path obeys
\bb
1 \leq \Tr[\rho B_0] \leq \|B_0\|_2 \leq \Tr[B_1] \leq \cdots \leq D\, .
\label{eq:budget}
\ee
\tcb{The inequalities after the first hold along every feasible path, so every feasible $B_j$ is
bounded; the constraints being non-strict, the feasible set of~\eqref{eq:dual} is closed as well, and the
maximum is attained. We bound the
primal variables by the same implication $S \geq \pm H \Rightarrow \Tr[S] \geq \|H\|_1$, applied now
to~\eqref{eq:chi}: a feasible chain has $S_i \geq 0$ and
\bbb
\Tr[S_{i - 1}] \;\leq\; \big\|S_{i - 1}^\Gamma\big\|_1 \;\leq\; \Tr[S_i]\qquad(0 \leq i \leq p)\, ,
\eee
so $\Tr[S_p] \leq c$ forces $0 \leq S_i \leq c\,\id$ at every index. The chains obeying this are a compact
set, nonempty for $c$ large enough, on which the minimum in~\eqref{eq:chi} is attained.}


Along an optimal path the traces start at no less than one, never decrease, and end at $D$, so the increments $\Tr[B_{j + 1}] - \Tr[B_j]$ are nonnegative and sum to at most $D - 1$.

Note that the value of the dual program depends only on the very first operator of the path, while the whole operator sequence after it is there just to certify that the first operator is legitimate. Therefore, the main idea is as follows: if we can keep a prefix of the path and replace the rest by something shorter, the value has not changed but the level has dropped. Proving the hierarchy collapses means proving that every long path can be rerouted in this way.

\tcb{We reroute at an index $k \in \{0,\ldots,p\}$, keeping $B_0,\ldots,B_k$ and replacing what follows by a short path from $B_k$ to $\id$. What is left is to find a suitable $k$, since we cannot choose it at will. We apply the next lemma with $C_j = B_{k + j}$. Its hypothesis is $B_{k + j + 1} - B_{k + j}^\Gamma \leq 2j\,\id$ for all $j \geq 1$, a bound on the slacks of the constraints from $B_{k + 1}$ onwards; we choose $k$ after stating the lemma.

It remains to reach $\id$ from $B_k$. The discarded tail may have any length, so the
replacement cannot be obtained by deleting steps from it; it has to be built anew, and its length has to be
controlled by the dimensions alone, since otherwise nothing has been gained. This is what the next lemma
supplies: given any feasible path from a Hermitian operator to $\id$, of whatever length, whose slacks obey this bound, it returns one
with the same endpoints and $n + 1$ transitions, where $n$ may be chosen in terms of $D$
alone.}

\begin{lemma}[(Tail compression)]\label{lem:compression}

Let $C_0,\ldots,C_m=\id$, with $m\geq1$, be Hermitian operators on a
bipartite Hilbert space of total dimension $D$. Assume that
\begin{alignat}{2}
-C_{j+1} \leq C_j^\Gamma&\leq C_{j+1}
  &\qquad &(0\leq j<m), \notag\\
C_{j+1}-C_j^\Gamma &\leq 2j\,\id
  &\qquad &(1\leq j<m).
\label{eq:slack-growth}
\end{alignat}
For any integer $n\geq0$ with $2n\geq D-\Tr[C_1]$, there are
Hermitian operators $C'_0,\ldots,C'_{n+1}$ with $C'_0=C_0$ and
$C'_{n+1}=\id$ satisfying
$$
-C'_{j+1}\leq (C'_j)^\Gamma\leq C'_{j+1}
\qquad (0\leq j\leq n).
$$
Each $C'_j$ is a real linear combination of the operators $C_i$ and $C_i^\Gamma$.
\end{lemma}

\begin{proof}

Set $C_j=\id$ for $j>m$ and $R_j\coloneqq C_{j+1}-C_j^\Gamma$
for $j\geq1$. Since $R_j=0$ for $j\geq m$,
\eqref{eq:slack-growth} extends to all $j\geq1$. Moreover, we have
\begin{equation}
\label{eq:tail-slacks}
\sum_{j\geq1}\Tr[R_j]=D-\Tr[C_1].
\end{equation}

We collect the slacks according to their parity, starting from each successive index:
\bb
P_j \coloneqq \sum_{\ell \geq 0}R_{j + 1 + 2\ell}\qquad(j \geq 0)\, .
\label{eq:corrections}
\ee
These sums are finite, because $R_j = 0$ for $j \geq m$. Telescoping the original path and transposing it alternately gives
\bb
\id - C_j = P_j + P_{j - 1}^\Gamma,\qquad P_{j - 1} - P_{j + 1} = R_j\qquad(j \geq 1)\, .
\label{eq:tail}
\ee
The $R_j$ are positive and their traces sum to $D-\Tr[C_1]$, so $0 \leq P_j \leq (D-\Tr[C_1])\id$. If $\Tr[C_1]=D$, every $P_j$ vanishes, and~\eqref{eq:tail} at $j = 1$ gives $C_1 = \id$, so the original first transition suffices.

For $n\geq1$, consider the linear interpolation with weight $(j-1)/n$
on $\id$ and the remaining weight on $C_j$. For $2\leq j\leq n$,
subtracting the partial transpose of the preceding term from the $j$th
term gives, by~\eqref{eq:tail},
\bbb
\frac{n-j+1}{n}R_{j-1}
+\frac{P_{j-2}^{\vphantom{\Gamma}}+P_{j-1}^{\Gamma}}{n}\, .
\eee 
Positivity of $P_{j-1}$ alone does not imply $P_{j-1}^\Gamma\geq0$.
Adding $P_j/n$ at step $j$ contributes $(P_j-P_{j-1}^\Gamma)/n$ to
this difference. The transposed term cancels, leaving the positive
operator $(P_{j-2}+P_j)/n$ in the second summand. We therefore keep
the endpoints and set
\bb
C'_j &\coloneqq \frac{n-j+1}{n}C_j+\frac{j-1}{n}\id+\frac{P_j}{n}\,,
\qquad 1\leq j\leq n\,.
\label{eq:interpolation}
\ee
\tcb{{Let us understand the above interpolation in detail. The three terms have three roles. The first follows the original path, $C_j$ at step $j$. The second is the identity, the only guaranteed positive ingredient and hence the resource
that `repairs' negativity; its weight $(j-1)/n$ climbs from $0$ to $(n-1)/n$ before the final step, so the new path is ``mostly the original one'' at the
start and ``mostly $\id$'' at the end. The third is the correction, present in exactly the amount needed to
cancel the transposed term.}}

The first transition is feasible because $C'_1 = C_1 + P_1/n$. At the intermediate transitions,
Eq.~\eqref{eq:tail} gives, for $2 \leq j \leq n$,
\bb
C'_j - (C'_{j - 1})^\Gamma &= \frac{n-j+1}{n}R_{j - 1} + \frac{P_{j - 2} + P_j}n\, ,\\
C'_j + (C'_{j - 1})^\Gamma &= \frac{n-j+1}{n}\big(C_j + C_{j - 1}^\Gamma\big) + \frac{2(j - 1)\id - R_{j - 1}}n\, .
\label{eq:slacks}
\ee
Every term on the right of the first line is positive. On the second line, the first term is positive by feasibility of the original path, and the second by~\eqref{eq:slack-growth} at index $j - 1$, which is the inequality $R_{j - 1}/n \leq 2(j-1)\id/n$. Moreover, \eqref{eq:tail} at $j = n$, partially transposed, gives
\bb
(C'_n)^\Gamma = \id - \frac{P_{n - 1}}n\, ,
\label{eq:endpoint}
\ee
so the last signed slacks are $P_{n - 1}/n$ and $2\id - P_{n - 1}/n$, both positive because $0 \leq P_{n - 1} \leq (D-\Tr[C_1])\id$ and $2n\geq D-\Tr[C_1]$. This also covers $n = 1$, when there are no intermediate transitions. Finally, all constructed operators are
linear combinations of the original ones and their partial transposes, because the same is true
of~\eqref{eq:corrections} and $\id = C_m$.
\end{proof}

We can now finish the proof of Theorem~\ref{thm:collapse}. Fix an optimal dual path for $p \geq p_\ast$, extend it by $B_t \coloneqq \id$ for $t\geq p+1$, and write $R_t \coloneqq B_{t + 1} - B_t^\Gamma \geq 0$ for $t \geq 1$, so that $R_t = 0$ for $t > p$ and $\Tr[R_t] = \Tr[B_{t + 1}] - \Tr[B_t]$. These increments sum to at most $D - 1$, so
\bb
\sum_{t \geq 1}\|R_t\|_\infty \;\leq\; \sum_{t \geq 1}\Tr[R_t] \;=\; D - \Tr[B_1] \;\leq\; D - 1\, ,
\label{eq:slack-sum}
\ee
where $\|\cdot\|_\infty$ denotes the operator norm, and the first inequality holds because $R_t \geq 0$. Let $k$ be the smallest nonnegative integer such that
\bb
R_{k + j} \leq 2j\,\id\qquad(j \geq 1)\, ;
\label{eq:first-index}
\ee
it exists and satisfies $k \leq p$, since $R_t = 0$ for $t > p$. We say that $t$ excludes $h$ if $0 \leq h < t$ and $\|R_t\|_\infty > 2(t - h)$. As $R_t \geq 0$, the bound $R_t \leq c\,\id$ is equivalent to $\|R_t\|_\infty \leq c$; hence every $h < k$, which fails~\eqref{eq:first-index} by the minimality of $k$, is excluded by some $t$. That $t$ satisfies $t \leq k$, since $t > k$ would give $\|R_t\|_\infty > 2(t - h) > 2(t - k)$, contradicting~\eqref{eq:first-index}. A fixed $t$ excludes only the $h$ with $1 \leq t - h < \|R_t\|_\infty/2$, at most $\|R_t\|_\infty/2$ of them, so counting the $k$ excluded integers $0,\ldots,k - 1$ gives
\bb
2k \;\leq\; \sum_{t = 1}^k\|R_t\|_\infty \;\leq\; \sum_{t = 1}^k\Tr[R_t]\, .
\label{eq:excluded}
\ee
We apply Lemma~\ref{lem:compression} with $C_j = B_{k + j}$, whose slacks are $R_{k + j}$, so that~\eqref{eq:slack-growth} is~\eqref{eq:first-index} and $\sum_{t>k}\Tr[R_t]=D-\Tr[B_{k+1}]$. If $\Tr[B_{k+1}]=D$, the single transition of the lemma gives the path $B_0,\ldots,B_k,\id$, of level $k \leq p_\ast$ by~\eqref{eq:slack-sum} and~\eqref{eq:excluded}. Otherwise we take $n = \lceil (D-\Tr[B_{k+1}])/2\rceil$; the compressed path has $k + n + 1$ transitions, and its level obeys
\bb
k + n \;&=\; \left\lceil k + \frac{D-\Tr[B_{k+1}]}2\right\rceil \;\leq\; \left\lceil\frac12\sum_{t \geq 1}\Tr[R_t]\right\rceil\\
&\leq\; \left\lceil\frac{D - 1}2\right\rceil \;=\; \left\lfloor\frac D2\right\rfloor \;=\; p_\ast\, .
\label{eq:level-count}
\ee
Padding with copies of $\id$ gives a certificate for $\chi_{p_\ast}$ with the original witness $B_0$, proving collapse. Indeed, the construction gives $\chi_p(\rho) \leq \chi_{p_\ast}(\rho)$ for every $p \geq p_\ast$, and monotonicity gives the reverse inequality. To identify the value with the upper hierarchy, take an optimal primal chain for $\chi_{p + 1}$. Its prefix is feasible for $\chi_p$, so
\bb
\chi_p(\rho) \leq \Tr[S_p] \leq \Tr[S_{p + 1}] = \chi_{p + 1}(\rho) = \chi_p(\rho)\, .
\label{eq:plateau}
\ee
The positive slack $S_{p + 1} - S_p^\Gamma$ consequently has zero trace. \tcb{A positive operator of zero trace vanishes, its eigenvalues being nonnegative with zero sum. Once $S_p^\Gamma \geq 0$, the chain continues at constant trace by $S_{j + 1} \coloneqq S_j^\Gamma$ for $j \geq p$, each step feasible because $S_j^\Gamma \geq 0$ is the same as $S_j^\Gamma \geq \pm S_j^\Gamma$, and each $S_{j + 1}^\Gamma = S_j$ positive in turn; collapse at one level therefore persists at every later one.} Hence $S_p^\Gamma = S_{p + 1} \geq 0$, and the same prefix is feasible for $\kappa_{p + 1}$, the program for $\chi_p$ with the additional constraint $S_p^\Gamma\geq0$, at cost $\chi_p$, proving~\eqref{eq:collapse} together with the operational characterisation of Ref.~\cite{LMR25}.

Finally, in the SM we carry the proof of Theorem~\ref{thm:collapse} over to ordered vector spaces (Sec.~\ref{sec:abstract-compression}). For Hermitian matrices, this shows that the proof uses no property of the partial transpose beyond its being a linear involution that preserves the identity, the trace and the Hilbert--Schmidt norm. 



\textbf{\em Discussion.}---Entanglement theory is full of regularised quantities and in desperate need of single-letter expressions to make it tractable. The scientific community has considered simplified settings to see if these regularisations can be done away with. For instance, Alice and Bob may be allowed to perform operations beyond LOCC, such as PPT operations. Despite such attempts, single-letter expressions have remained elusive to date, in particular for the LOCC cost~\eqref{eq:locc-cost}, which is the regularised entanglement of formation~\cite{asympform}. The entanglement of formation is not additive~\cite{Shor2004,Hastings2009}, and no single-letter expression for the LOCC cost is known. In the PPT setting, Theorem~\ref{thm:collapse} provides one: the exact PPT entanglement cost is given by a single semidefinite program of predetermined size. To our knowledge, it is the first operational measure with such a formula.

The formula is remarkable in its own right, and its implications extend beyond the PPT setting, for example to the exact cost under LOCC. That cost is given by the regularised Schmidt number~\cite{nielsen99}, and the regularisation cannot be removed because the Schmidt number of a single copy is submultiplicative~\cite{horodecki00,Yue2019}. As a result, the exact LOCC cost is known only for a handful of states~\cite{Yue2019,SenLami2026Flower}, and lower bounds on it are valuable. This is where the importance of our Theorem~\ref{thm:collapse} lies. The exact PPT cost is by definition a lower bound on the exact LOCC cost, and by Theorem~\ref{thm:collapse} this lower bound equals $\log_2\chi_{p_\ast}(\rho)$ ebits per copy, the value of a single finite semidefinite program. Within the PPT setting, the theorem also extends the characterisation~\eqref{eq:zero-binegativity} of Audenaert, Plenio and Eisert~\cite{audenaert2003entanglement} from states of zero binegativity to all bipartite states: $\log_2\chi_p(\rho)$ equals $E_N(\rho)$ at $p=0$ and $E_{c,\ppt}^{\mathrm{exact}}(\rho)$ for every $p\geq p_\ast$.


Every two-qubit state has zero binegativity~\cite{Satoshi},
so logarithmic negativity gives the exact PPT
cost~\cite{audenaert2003entanglement}. In higher dimensions,
logarithmic negativity, given by $\log_2\Tr|\rho^\Gamma|$,
can underestimate the cost~\cite{LMR25}, but it can easily be proved that the same operation
$\ast\mapsto|\ast^\Gamma|$ also provides an upper bound
when repeated starting from $\rho$ until the resulting operator is
PPT~\cite[Lemma~4]{Brinster2026}. In Theorem~\ref{thm:greedy} of the
SM, we prove that every finite-dimensional bipartite state $\rho$ reaches such a PPT operator within an explicit number of steps, thus giving a simple (negativity-like) upper bound on the cost.



 \medskip
\textbf{\em Acknowledgements.}--- S.S.\ and L.L.\ acknowledge financial support from the European Union (ERC StG ETQO, Grant Agreement no.\ 101165230). The authors acknowledge the use of ChatGPT 5.6 Sol and ChatGPT 6 Astra for assistance in deriving results. The authors take full responsibility for the accuracy and integrity of this work.

\medskip
\emph{Note added.}---During the preparation of this work, we became
aware of related work by Song et al.~\cite{Song2026Qutrit},
who obtain a single-letter formula for the exact PPT entanglement
cost of all two-qutrit states. Our work, however, is not restricted to two-qutrit systems only and applies to any arbitrary finite dimensional system.

\bibliographystyle{linked-unsrt}
\bibliography{references}

\let\addcontentsline\oldaddcontentsline
\clearpage
\hypersetup{pageanchor=false}
\fakepart{Supplemental Material}
\onecolumngrid
\begin{center}
\vspace*{\baselineskip}
{\textbf{\large Supplemental Material}}\\
\end{center}
\renewcommand{\theequation}{S\arabic{equation}}
\renewcommand{\thethm}{S\arabic{thm}}
\renewcommand{\thefigure}{S\arabic{figure}}
\setcounter{page}{1}
\setcounter{secnumdepth}{2}
\setcounter{tocdepth}{2}
\makeatletter
\let\l@title\@gobbletwo
\let\l@abstract\@gobbletwo
\makeatother
\hypersetup{linktoc=all}
\begingroup\color{Blues5seq5}
\tableofcontents
\endgroup
\medskip

\tcb{ Section~\ref{sec:iterated} proves Theorem~\ref{thm:greedy}: the iteration $X_0 = \rho$, $X_{p + 1} = \big|X_p^\Gamma\big|$ reaches a PPT operator within $\pazocal{O}(D^{3/2}\log D)$ steps.}

Section~\ref{sec:abstract-compression} \tcb{carries the proof of Theorem~\ref{thm:collapse} over to an ordered vector space}.

\medskip

\section{Iterated negativities and finite termination}\label{sec:iterated}

Ishizaka showed that $\big|\rho^\Gamma\big|^\Gamma \geq 0$ for every two-qubit state $\rho$~\cite{Satoshi}. By the result of Audenaert, Plenio and Eisert~\cite{audenaert2003entanglement} stated in~\eqref{eq:zero-binegativity}, the exact PPT entanglement cost of every two-qubit state therefore equals its logarithmic negativity $\log_2\big\|\rho^\Gamma\big\|_1$.

In higher dimensions, $\big|\rho^\Gamma\big|$ can have a negative partial transpose. Does repeating the operation $X \mapsto \big|X^\Gamma\big|$ eventually produce a PPT operator? Define
\bb
X_0 \coloneqq \rho\, ,\qquad X_{p + 1} \coloneqq \big|X_p^\Gamma\big|\, .
\label{eq:greedy}
\ee
If $X_p$ is PPT for some $p \geq 1$, then $\log_2\Tr[X_p]$ is an upper bound on $E_{c,\ppt}^{\mathrm{exact}}(\rho)$~\cite[Lemma~4]{Brinster2026}. More precisely, for every integer $p\geq1$, we have~\cite[Lemma~4]{Brinster2026}
\bb
X_p\ \text{is PPT}
\quad\Longrightarrow\quad
E_{c,\ppt}^{\mathrm{exact}}(\rho)\leq \log_2\Tr[X_p]\, .
\label{eq:greedy-upper-bound}
\ee
This can be easily proved as follows: if $X_p$ is PPT, the operators $X_1,\ldots,X_p$ are feasible
for $\kappa_p$~\cite{LMR25}, and the upper bound of Ref.~\cite{LMR25} applies; see also~\cite[Lemma~4]{Brinster2026}.

We show that every state reaches such a PPT iterate within a number of steps bounded in terms of the dimensions alone.

\begin{thm}[(Finite termination)]\label{thm:greedy}
For every state $\rho$ with $d=\min\{d_A,d_B\}\geq2$, the operator $X_p$ is PPT for every integer $p$ satisfying
\bb
p\geq \left\lceil 3D^{3/2}\ln(32D)\right\rceil\, .
\label{eq:greedy-level}
\ee
\end{thm}

\begin{proof}
\emph{Finite termination for a fixed state.} Taking an absolute value preserves the Hilbert--Schmidt norm, and so does partial transposition, so $\|X_p\|_2 = \|\rho\|_2$ at every step. Applying $\Gamma^p$ to the $p$th iterate removes the alternating partial transpose. Write $Y_p \coloneqq \Gamma^p(X_p)$ and $N_p \coloneqq \big(X_p^\Gamma\big)_-$, where $H_- \coloneqq (|H| - H)/2$ denotes the negative part. Then
\bb
Y_{p + 1} - Y_p = 2\Gamma^{p + 1}(N_p)\, ,\qquad \Tr[X_{p+1}]-\Tr[X_p] = 2\Tr[N_p]\, .
\label{eq:unfolded}
\ee
The trace is nondecreasing and bounded by $\sqrt D\,\|\rho\|_2$, so its increments are summable. By~\eqref{eq:unfolded}, the length of each step is at most the corresponding trace increment, since $\|N_p\|_2 \leq \Tr[N_p]$. Thus $Y_p$ converges to an operator $Z$. The even $Y_p$ are positive and the odd $Y_p$ have positive partial transpose, so $Z$ is PPT.

Write $s_p \coloneqq \Tr[Z] - \Tr[X_p]$ for the trace still to be gained, which tends to zero. To show that it reaches zero after finitely many steps, we relate each positive $s_p$ to an eigenvalue of $Z$ or $Z^\Gamma$. Norm conservation gives, for every PPT operator $W$,
\bb
\|Y_j - W\|_2^2 - \|Y_{j + 1} - W\|_2^2 = 4\Tr\big[\Gamma^{j + 1}(W)N_j\big] \geq 0\, .
\label{eq:distance}
\ee
Indeed, expanding the squares cancels the two norms of the iterates, and the remaining term follows from~\eqref{eq:unfolded}. Taking $W = Z$, write $r_j \coloneqq \|Z - Y_j\|_2$ and $T_j \coloneqq \Gamma^{j + 1}(Z)$. Then $r_j$ is nonincreasing, and summing~\eqref{eq:distance} and the trace increments from step $p$ onwards gives
\bb
r_p^2 = 4\sum_{j \geq p}\Tr[T_jN_j]\, ,\qquad s_p = 2\sum_{j \geq p}\Tr[N_j]\, .
\label{eq:remainders}
\ee

Fix $p$ with $s_p > 0$, and let $\lambda$ be the largest eigenvalue of $Z$ or $Z^\Gamma$ that is at most $4s_p$, taking $\lambda = 0$ if there is none. Since $T_j$ is one of these two operators, every eigenvalue $x$ of $T_j$ satisfies either $x \leq \lambda$ or $x > 4s_p$, and in the latter case $x \leq x^2/(4s_p)$. Hence $T_j \leq \lambda\id + T_j^2/(4s_p)$. To use this inequality in~\eqref{eq:remainders}, it suffices to test it on the negative eigenvectors of $X_j^\Gamma$. If $X_j^\Gamma v = -av$ with $a > 0$ and $\|v\| = 1$, positivity of $T_j$ gives
\bbb
\|T_jv\| \;\leq\; \|(T_j + a\id)v\| \;=\; \big\|\big(T_j - X_j^\Gamma\big)v\big\| \;\leq\; \big\|T_j - X_j^\Gamma\big\|_2 \;=\; r_j\, ,
\eee
the last equality because $T_j - X_j^\Gamma = \Gamma^{j + 1}(Z - Y_j)$. Weighting these eigenvectors by their eigenvalues in $N_j$, we obtain $\Tr[T_jN_j] \leq \big(\lambda + r_j^2/(4s_p)\big)\Tr[N_j]$. Since $r_j \leq r_p$ for $j \geq p$, substitution into~\eqref{eq:remainders} gives $r_p^2 \leq 2s_p\lambda + r_p^2/2$, that is, $r_p^2 \leq 4s_p\lambda$. Finally, $s_p = \Tr[Z - Y_p] \leq \sqrt D\,r_p$, so
\bb
\frac\lambda4 \;\leq\; s_p \;\leq\; 4D\lambda\, .
\label{eq:window}
\ee
In particular, $\lambda > 0$ whenever $s_p > 0$. The smallest positive eigenvalue in the combined spectra of $Z$ and $Z^\Gamma$ exists because $Z \geq 0$ and $\Tr[Z] \geq \Tr[\rho] = 1$. Every positive $s_p$ is therefore at least one quarter of this eigenvalue. Since $s_p \to 0$, some finite $p$ has $s_p = 0$, and the second identity in~\eqref{eq:remainders} then gives $N_p = 0$, so $X_p$ is PPT. From that step onwards $X_{p + 1} = X_p^\Gamma$ and $X_{p + 2} = X_p$.

\emph{A stopping level uniform over states.} This eigenvalue depends on the input. To count the steps using only the dimensions, we return to~\eqref{eq:window}. There are at most $2D$ intervals $[\lambda/4,4D\lambda]$, each spanning a factor $16D$, and it remains to bound how many steps the trace remainder can spend in any one of them. We do this by showing that $s_{p + 1} \leq qs_p$ for a constant $0 < q < 1$ depending only on $d$ and $D$.

The idea is to correct the negative part $N_p$ at the current step and to use~\eqref{eq:distance} to bound how far the limit can be from the corrected operator. For this purpose, every $N \geq 0$ admits a PPT majorant $R \geq N$ satisfying
\bb
\Tr[R] \leq d\Tr[N]\, ,\qquad \|R\|_2 \leq \sqrt{\frac{2d}{d + 1}}\,\Tr[N]\, .
\label{eq:majorant}
\ee
To see this first for $N = |\psi\rangle\langle\psi|$, write a unit vector in Schmidt form as $|\psi\rangle = \sum_{i = 1}^d\sqrt{u_i}\,|ii\rangle$, padding the Schmidt probabilities with zeros if necessary, and take
\bb
R \coloneqq \frac d{d + 1}\left(|\psi\rangle\langle\psi| + \sum_{i \ne j}\sqrt{u_iu_j}\,|ij\rangle\langle ij|\right) + \frac1{d + 1}\sum_i |ii\rangle\langle ii|\, .
\label{eq:majorant-construction}
\ee
The last sum projects onto the correlated Schmidt subspace and hence dominates $|\psi\rangle\langle\psi|$, proving $R \geq N$. Under partial transpose in the Schmidt basis, each pair $|ij\rangle,|ji\rangle$ with $i < j$ gives a two-dimensional block whose four entries are all $d\sqrt{u_iu_j}/(d + 1)$. These blocks are positive, as are the remaining diagonal entries. Direct calculation gives
\bb
\Tr[R] &= \frac d{d + 1}\left[\left(\sumno_i\sqrt{u_i}\right)^2 + 1\right] \leq d\, ,\\
\|R\|_2^2 &= \frac{d^2\left(2 - \sum_iu_i^2\right) + 3d}{(d + 1)^2} \leq \frac{2d}{d + 1}\, ,
\label{eq:majorant-estimates}
\ee
where $\big(\sum_i\sqrt{u_i}\big)^2 \leq d$ and $\sum_iu_i^2 \geq 1/d$. A spectral decomposition of $N$ and the triangle inequality extend~\eqref{eq:majorant} to arbitrary $N \geq 0$.

Apply this construction to $N_p$ and set $W \coloneqq Y_p + \Gamma^{p + 1}(R)$. The operators $W$ and $W^\Gamma$ are, in some order, $X_p + R^\Gamma$ and $X_p^\Gamma + R$, both positive since $R$ is PPT and $R \geq N_p$. Thus~\eqref{eq:distance} applies to this $W$, and taking the limit gives $\|Z - W\|_2 \leq \|Y_p - W\|_2 = \|R\|_2$. Since $\Tr[W - Y_p] = \Tr[R]$, it follows that
\bb
s_p \;=\; \Tr[Z - W] + \Tr[R] \;\leq\; \sqrt D\,\|R\|_2 + \Tr[R] \;\leq\; \frac{d + \sqrt{2dD/(d + 1)}}2\,(\Tr[X_{p+1}]-\Tr[X_p])\, .
\label{eq:contraction}
\ee
As $\Tr[X_{p+1}]-\Tr[X_p] = s_p - s_{p + 1}$, this is the contraction $s_{p + 1} \leq qs_p$ with
\bbb
q \coloneqq 1 - \frac2{d + \sqrt{2dD/(d + 1)}} \in (0,1)\, .
\eee

Suppose $s_p$ and $s_{p + k}$ both lie in the interval associated with $\lambda > 0$. Then
\bbb
\frac\lambda4 \;\leq\; s_{p + k} \;\leq\; q^ks_p \;\leq\; q^k\,4D\lambda\, .
\eee
The eigenvalue cancels, giving $k \leq \ln(16D)/(-\ln q)$. Thus each interval contains at most $1 + \lfloor\ln(16D)/(-\ln q)\rfloor$ indices, however close it lies to zero. Every index before the first PPT iterate has $s_p > 0$ and is covered by one of the at most $2D$ intervals. Multiplying the two counts shows that the first PPT iterate has index at most
\bb
2D\left(1 + \left\lfloor\frac{\ln(16D)}{-\ln q}\right\rfloor\right) \;\leq\; 2D + (1 + \sqrt2)D^{3/2}\ln(16D) \;\leq\; 3D^{3/2}\ln(32D)\, .
\label{eq:greedy-count}
\ee
The first inequality holds because $-\ln q \geq 1 - q$, $d \leq \sqrt D$ and $2d/(d + 1) < 2$ give $2/(-\ln q) \leq d + \sqrt{2dD/(d + 1)} \leq (1 + \sqrt2)\sqrt D$, and the second because $1 + \sqrt2 < 3$ and $2D \leq 3D^{3/2}\ln2$. Since every iterate following a PPT iterate is PPT, $X_p$ is PPT for every $p$ satisfying~\eqref{eq:greedy-level}.
\end{proof}

\section{Compression in an ordered vector space}\label{sec:abstract-compression}

The proof of Theorem~\ref{thm:collapse} uses less structure than partial transposition. Its
essential ingredients are a positive functional that bounds the total increase along a path and an order unit
that controls the size of positive corrections. We formulate the resulting principle for a fixed
ordered vector space.

\begin{thm}[(Compression at fixed ordered geometry)]\label{thm:abstract-compression}
Let $V$ be a finite-dimensional real vector space, let $K \subset V$ be a closed, pointed, generating convex
cone, and write $X \geq Y$ when $X - Y \in K$. Fix an order unit $e \in \operatorname{int}K$, a linear
functional $\tau$ that is strictly positive on $K \setminus\{0\}$, and a linear involution $\Gamma$ satisfying
$\Gamma(e) = e$ and $\tau \circ \Gamma = \tau$. There exists a finite constant $\beta > 0$ such that
\bb
X \leq \beta\tau(X)e\qquad(X \in K)\, .
\label{eq:abstract-order-bounds}
\ee
For any such $\beta$, every finite path
$C_0,\ldots,C_m$ with $m\geq1$ satisfying $C_m = e$ and $C_{j + 1} \geq \pm\Gamma(C_j)$ can be replaced, with
the same endpoints, by a path of at most
\bb
\left\lceil\frac{\beta\big(\tau(e) - \tau(C_1)\big)}2\right\rceil + 1
\label{eq:abstract-horizon}
\ee
transitions. Every member of the replacement path is a real linear combination of the original $C_j$ and
$\Gamma(C_j)$.
\end{thm}

\begin{proof}
The base $\{X \in K:\tau(X) = 1\}$ is compact. Indeed, an unbounded sequence in this base, divided by its norm
and restricted to a convergent subsequence, would give a nonzero element of $K$ on which $\tau$ vanishes.
Since $e$ is interior, bounded sets are bounded above by scalar multiples of $e$: if
$e - Z \in K$ whenever $\|Z\|\leq r$, and $\|X\|\leq M$ with $M,r>0$, then $\|r X/M\|\leq r$ and
hence $(M/r)e-X \in K$. Applying this to the base gives~\eqref{eq:abstract-order-bounds} by
homogeneity.

Beyond the order itself, the only matrix inequality in the proof of Theorem~\ref{thm:collapse} is $R \leq \Tr[R]\id$ for $R \geq 0$, used for $P_j$ in Lemma~\ref{lem:compression} and for $\|R_t\|_\infty$ in the choice of $k$; \eqref{eq:abstract-order-bounds} replaces it. Extend the path by $C_j \coloneqq e$ for $j > m$, and set $R_j \coloneqq C_{j + 1} - \Gamma(C_j) \geq 0$ for $j \geq 1$. Since $\tau \circ \Gamma = \tau$, the sum telescopes to $\sum_{j \geq 1}\tau(R_j) = \tau(e) - \tau(C_1)$. For $R \geq 0$ write $c(R) \coloneqq \min\{c \geq 0 : R \leq c\,e\}$, which exists because $K$ is closed; then $R \leq c\,e$ holds exactly when $c \geq c(R)$, and $c(R) \leq \beta\tau(R)$ by~\eqref{eq:abstract-order-bounds}.

First, Lemma~\ref{lem:compression} holds with $e$ in place of $\id$, $\Gamma(C_j)$ in place of $C_j^\Gamma$, and the condition $2n\geq D-\Tr[C_1]$ replaced by $2n\geq\beta(\tau(e)-\tau(C_1))$. Indeed, $P_j$ defined by~\eqref{eq:corrections} satisfies~\eqref{eq:tail}, whose derivation uses only linearity, $\Gamma^2 = \mathrm{id}$ and $\Gamma(e) = e$, and $0 \leq P_j \leq \beta\tau(P_j)e \leq \beta(\tau(e)-\tau(C_1))e$. The slacks~\eqref{eq:slacks} and~\eqref{eq:endpoint} are unchanged, the final ones being $P_{n - 1}/n$ and $2e - P_{n - 1}/n$. If $\tau(C_1)=\tau(e)$, every $R_j$ vanishes, because $\tau$ is strictly positive on $K \setminus\{0\}$.

Second, the choice of $k$ in the proof of Theorem~\ref{thm:collapse}, applied to $C_0,\ldots,C_m$ in place of $B_0,\ldots,B_{p + 1}$, goes through with $\|R_t\|_\infty$ replaced by $c(R_t)$ and $\Tr$ by $\tau$, and~\eqref{eq:excluded} becomes
\bbb
2k \;\leq\; \sum_{t = 1}^k c(R_t) \;\leq\; \beta\sum_{t = 1}^k\tau(R_t)\, .
\eee
With $n=\lceil\beta(\tau(e)-\tau(C_{k+1}))/2\rceil$, the replacement path has $k + n + 1$ transitions, and
\bbb
k + n \;\leq\; \left\lceil\frac\beta2\sum_{t \geq 1}\tau(R_t)\right\rceil \;=\; \left\lceil\frac{\beta\big(\tau(e) - \tau(C_1)\big)}2\right\rceil\, .
\eee
If $\tau(C_{k+1})=\tau(e)$, the path $C_0,\ldots,C_k,e$ has $k + 1$ transitions, which obey the same bound. The linear-combination form follows as in Lemma~\ref{lem:compression}.
\end{proof}


\subsection{The role of the partial transpose}
For the cone of positive semidefinite Hermitian $D \times D$ matrices, with $e = \id$ and $\tau = \Tr$, every $X \geq 0$ satisfies $X \leq \Tr[X]\id$. Hence~\eqref{eq:abstract-order-bounds} holds with $\beta = 1$ for every unital, trace-preserving linear involution $\Gamma$ of the Hermitian matrices, and Theorem~\ref{thm:abstract-compression} gives replacement paths with at most $\lceil(D - \Tr[C_1])/2\rceil + 1$ transitions. If $\Gamma$ is in addition an isometry for the Hilbert--Schmidt norm, and hence self-adjoint, \eqref{eq:dual} and~\eqref{eq:budget} hold as written, and the hierarchy~\eqref{eq:chi} with $\Gamma$ in place of the partial transpose collapses at level $\lfloor D/2\rfloor$, as in the proof of Theorem~\ref{thm:collapse}. That proof therefore uses no property of partial transposition beyond these.

\end{document}